\documentclass[longauth]{aa}  
\usepackage{ulem}
\usepackage{graphicx}
\usepackage{txfonts}
\usepackage[colorlinks=true,linkcolor=red,anchorcolor=blue,citecolor=blue,filecolor=black,menucolor=black,runcolor=black,urlcolor=blue]{hyperref}

\newcommand{\arcs}{$\arcsec$\xspace}

\begin{document} 
 
    \title{Spatially resolved H$\alpha$ and ionizing photon production efficiency in the lensed galaxy MACS1149-JD1 at a redshift of 9.11} 

   \author{J. \'Alvarez-M\'arquez\inst{\ref{inst:CAB}} \and L. Colina\inst{\ref{inst:CAB}} \and A. Crespo G\'omez\inst{\ref{inst:CAB}} \and P. Rinaldi\inst{\ref{inst:Groningen}} \and J. Melinder\inst{\ref{inst:Stockholm}} \and G. {\"O}stlin\inst{\ref{inst:Stockholm}} \and M. Annunziatella\inst{\ref{inst:CAB}} \and A. Labiano\inst{\ref{inst:Telespazio}} \and A. Bik\inst{\ref{inst:Stockholm}} \and S. Bosman\inst{\ref{inst:HeidelbergUni},\ref{inst:MPIA}}  \and T.R. Greve\inst{\ref{inst:DTU},\ref{inst:DAWN},\ref{inst:UCL}} 
   \and G. Wright\inst{\ref{inst:UKATC}} \and A. Alonso-Herrero\inst{\ref{inst:CAB-ESAC}} \and L. Boogaard\inst{\ref{inst:MPIA}} \and R. Azollini\inst{\ref{inst:CAB}, \ref{inst:Dublin}} \and K.I. Caputi\inst{\ref{inst:Groningen},\ref{inst:DAWN}} \and L. Costantin\inst{\ref{inst:CAB}} \and A. Eckart\inst{\ref{inst:Köln}} \and M. Garc\'ia-Mar\'in\inst{\ref{inst:ESA}} \and S. Gillman\inst{\ref{inst:DTU}, \ref{inst:DAWN}} \and J. Hjorth\inst{\ref{inst:DARK}} \and E. Iani\inst{\ref{inst:Groningen}} \and O. Ilbert\inst{\ref{inst:LAM}} \and I. Jermann\inst{\ref{inst:DTU}, \ref{inst:DAWN}} \and D. Langeroodi\inst{\ref{inst:DARK}} \and R. Meyer\inst{\ref{inst:geneve}} \and  F. Pei{\ss}ker\inst{\ref{inst:Köln}} \and P. P\'erez-Gonz\'alez\inst{\ref{inst:CAB}} \and J.P. Pye\inst{\ref{inst:Leicester}} \and T. Tikkanen\inst{\ref{inst:Leicester}} \and M. Topinka\inst{\ref{inst:Dublin}} \and P. van der Werf\inst{\ref{inst:Leiden}} \and F. Walter\inst{\ref{inst:MPIA}} \and Th. Henning\inst{\ref{inst:MPIA}} \and T. Ray\inst{\ref{inst:Dublin}}} 

   \institute{Centro de Astrobiolog\'{\i}a (CAB), CSIC-INTA, Ctra. de Ajalvir km 4, Torrej\'on de Ardoz, E-28850, Madrid, Spain\\  \email{jalvarez@cab.inta-csic.es} \label{inst:CAB}
    \and Kapteyn Astronomical Institute, University of Groningen, P.O. Box 800, 9700 AV Groningen, The Netherlands \label{inst:Groningen}
    \and Department of Astronomy, Stockholm University, Oscar Klein Centre, AlbaNova University Centre, 106 91 Stockholm, Sweden \label{inst:Stockholm}
    \and Telespazio UK for the European Space Agency, ESAC, Camino Bajo del Castillo s/n, 28692 Villanueva de la Ca\~{n}ada, Spain \label{inst:Telespazio}
    \and Institute for Theoretical Physics, Heidelberg University, Philosophenweg 12, D-69120, Heidelberg, Germany \label{inst:HeidelbergUni}
    \and Max-Planck-Institut f\"ur Astronomie, K\"onigstuhl 17, 69117 Heidelberg, Germany\label{inst:MPIA}
    \and DTU Space, Technical University of Denmark, Elektrovej 327, 2800 Kgs. Lyngby, Denmark \label{inst:DTU}
    \and Cosmic Dawn Centre (DAWN), Copenhagen, Denmark \label{inst:DAWN}
    \and Department of Physics and Astronomy, University College London, Gower Place, London WC1E 6BT, UK \label{inst:UCL}
    \and UK Astronomy Technology Centre, Royal Observatory Edinburgh, Blackford Hill, Edinburgh EH9 3HJ, UK \label{inst:UKATC} 
    \and Centro de Astrobiolog\'ia (CAB), CSIC-INTA, Camino Viejo del Castillo s/n, 28692 Villanueva de la Ca\~{n}ada, Madrid, Spain \label{inst:CAB-ESAC}
    \and Dublin Institute for Advanced Studies, Astronomy \& Astrophysics Section, 31 Fitzwilliam Place, Dublin 2, Ireland \label{inst:Dublin}
    \and I. Physikalisches Institut der Universit\"at zu K\"oln, Z\"ulpicher Str. 77, 50937 K\"oln, Germany \label{inst:Köln}
    \and European Space Agency, Space Telescope Science Institute, Baltimore, Maryland, USA \label{inst:ESA} 
    \and DARK, Niels Bohr Institute, University of Copenhagen, Jagtvej 128, 2200 Copenhagen, Denmark \label{inst:DARK} 
     \and Aix Marseille Universit\'e, CNRS, LAM (Laboratoire d’Astrophysique de Marseille) UMR 7326, 13388, Marseille, France \label{inst:LAM} 
    \and Department of Astronomy, University of Geneva, Chemin Pegasi 51, 1290 Versoix, Switzerland\label{inst:geneve} 
    \and School of Physics \& Astronomy, Space Research Centre, Space Park Leicester, University of Leicester, 92 Corporation Road, Leicester, LE4 5SP, UK \label{inst:Leicester}   
    \and Dublin Institute for Advanced Studies, Astronomy \& Astrophysics Section, 31 Fitzwilliam Place, Dublin 2, Ireland \label{inst:Dublin}
    \and Leiden Observatory, Leiden University, PO Box 9513, 2300 RA Leiden, The Netherlands \label{inst:Leiden}
   }

   \date{Received ; accepted}

\abstract
{We present MIRI/JWST medium-resolution spectroscopy (MRS) and imaging (MIRIM) of the lensed galaxy MACS1149-JD1 at a redshift of $z$\,=\,9.1092$\pm$0.0002, when the Universe was about 530 Myr old. We detect, for the first time, spatially resolved H$\alpha$ emission in a galaxy at a redshift above nine. The structure of the H$\alpha$ emitting gas consists of two clumps, S and N, carrying about $60\%$ and $40\%$ of the total flux, respectively. The total H$\alpha$ luminosity implies an instantaneous star-formation rate in the range of 3.2\,$\pm$\,0.3 and 5.3\,$\pm$\,0.4\,$M_{\odot}$\,yr$^{-1}$ for sub-solar and solar metallicities. The ionizing photon production efficiency, $\log(\zeta_\mathrm{ion})$, shows a spatially resolved structure with values of 25.55\,$\pm$\,0.03; 25.47\,$\pm$\,0.03; and 25.91\,$\pm$\,0.09\,Hz\,erg$^{-1}$ for the integrated galaxy and clumps S and N, respectively. The H$\alpha$ rest-frame equivalent width, EW$_{0}$\,(H$\alpha$), is 726$^{+660}_{-182}$\,$\AA$ for the integrated galaxy, but it presents extreme values of 531$^{+300}_{-96}\AA$ and $\geq$1951\,$\AA$ for clumps S and N, respectively. The spatially resolved ionizing photon production efficiency is within the range of values measured in galaxies at a redshift above six and well above the canonical value (25.2\,$\pm$\,0.1\,Hz\,erg$^{-1}$). The EW$_{0}$\,(H$\alpha$) is a factor of two lower than the predicted value at $z$\,=\,9.11 based on the extrapolation of the evolution of the EW$_{0}$\,(H$\alpha$) with redshifts, $\propto$ (1+z)$^{2.1}$, including galaxies detected with JWST. The extreme difference of the EW$_{0}$\,(H$\alpha$) for clumps S and N indicates the presence of a recent (<5 Myr) stellar burst in clump N and a star formation over a larger period of time (e.g., $\sim$50 Myr) in clump S. The different ages of the stellar population place MACS1149-JD1 and clumps N and S at different locations in the log($\zeta_\mathrm{ion}$) to EW$_{0}$\,(H$\alpha$) plane and above the main relation defined from intermediate- and high-redshift (z=3-7) galaxies detected with JWST. Finally, clump S and N show very different H$\alpha$ kinematics, with velocity dispersions of 56\,$\pm$\,4\,km\,s$^{-1}$ and 113\,$\pm$\,33\,km\,s$^{-1}$, likely indicating the presence of outflows or increased turbulence in clump N. The dynamical mass $M_\mathrm{dyn}$= (2.4\,$\pm$\,0.5)\,$\times$\,10$^{9}$\,$M_{\odot}$, obtained from the size of the integrated H$\alpha$ ionized nebulae and its velocity dispersion, is within the range previously measured with the spatially resolved [OIII]88$\mu$m line.
}

\titlerunning{Lensed MACS1149-JD1 galaxy at $z$\,=\,9.11}
\maketitle

\section{Introduction}\label{Sect:intro}

The quest for the first galaxies that reionized the Universe at redshifts six to 10 during the so-called Epoch of Reionization (EoR) is one of the main goals of the \textit{James Webb Space Telescope} (JWST; \citealt{Gardner+23}). Already in its first year of operation, JWST has pushed the limits of our knowledge of galaxy formation in the early universe well into the EoR. In addition to photometrically identified galaxy candidates up to $z$\,$\sim$\,16 \citep{Adams+23,Donnan+23,Finkelstein+23,Harikane+23a,Perez-Gonzalez+23b}, a large number of galaxies have already been spectroscopically confirmed at redshifts above six and up to a redshift of 13.2 \citep{Arrabal-Haro+23, Boyett+23, Bunker+23, Curtis-Lake+23, Fujimoto+23, Matthee+23, Williams+23}. While most sources at those redshifts are star-forming galaxies (SFGs), active galactic nuclei (AGNs) less massive than the already known $z$\,$>$\,6 quasars \citep{Yang+23} have also been detected through the presence of broad H$\beta$ lines \citep{Harikane+23b,Kocevski+23,Larson+23,Maiolino+23b, Maiolino+23a}. The study of the nature of these EoR sources and their contribution to the reionization of the Universe is now possible thanks to JWST. Deep rest-frame ultraviolet imaging combined with the spectroscopy of hydrogen Balmer emission lines provides an opportunity to directly measure the ionizing flux and therefore establish accurate measurements of key quantities, such as the instantaneous star-formation rate (SFR), the UV luminosity function, the ionizing photon production efficiency ($\zeta_\mathrm{ion}$), and the escape fraction ($f_\mathrm{esc}$).  

\begin{figure*}
\centering
   \includegraphics[width=\linewidth]{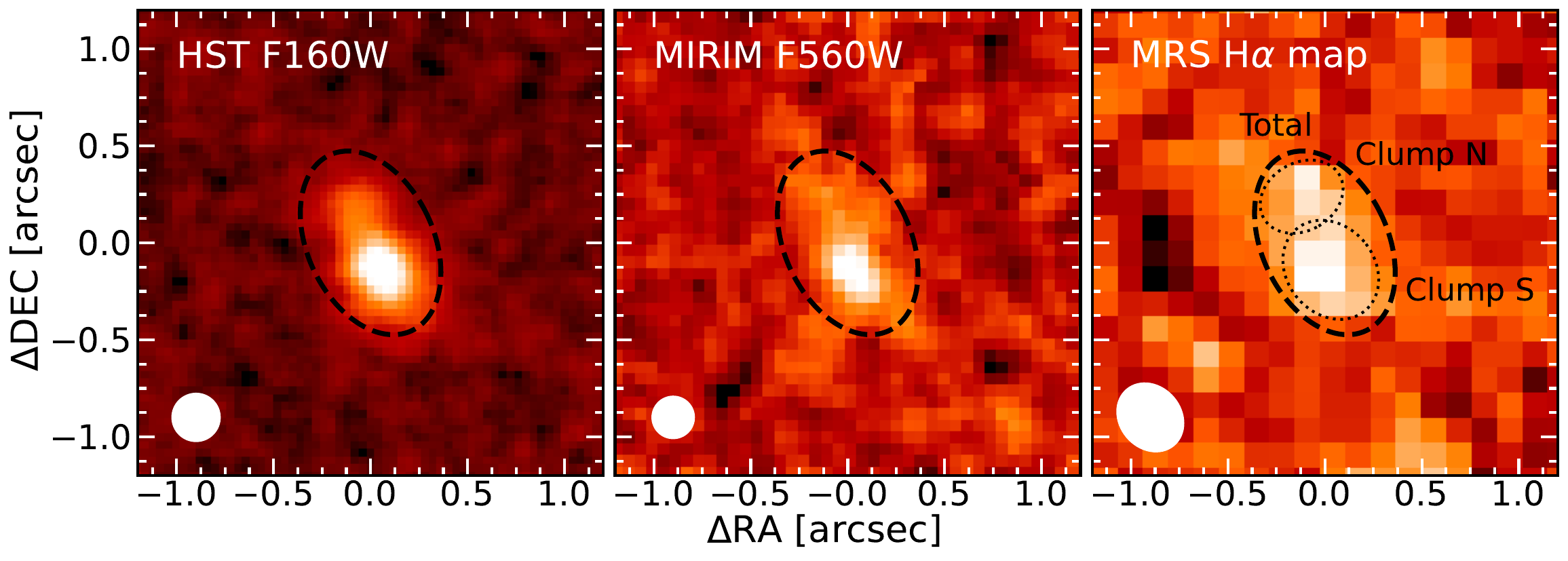}
      \caption{Images of MACS1149-JD1 presenting the observed stellar and ionized gas structure. From left to right: Archival HST WFPC3/F160W from Hubble Advanced Product Multi-Visit Mosaic (HAP-MVM) program, MIRIM F560W image, and MRS H$\alpha$ line map. The H$\alpha$ line map was generated by integrating H$\alpha$ line emission in the velocity range, -150\,<\,$v$\,[km\,s$^{-1}$]\,<\,150. The origin of the image corresponds to (RA\,[deg], DEC\,[deg]) of (177.389945, +22.412722). The black dashed ellipse indicates the aperture used to perform the MIRIM F560W photometry and the MRS 1D spectral extraction. Dotted black ellipses identify the H$\alpha$ emitting regions referred to as clump N and S, which are spatially coincident with the two emitting regions in the HST F160W image. Moreover, clump N corresponds to clump C1, and clump S includes clumps C2 and C3 as well as the galaxy component (G) following \cite{Bradac+23} nomenclature. The white area represents the spatial resolution (PSF FWHM) of each observation. MACS1149-JD1 shows an elongated structure due to the lens magnification of cluster MACS J11491+2223 \citep{Zheng+12}. }
         \label{fig:HaLineMap}
\end{figure*}

Prior to JWST, MACS1149-JD1 was the highest redshift galaxy spectroscopically confirmed using emission lines. MACS1149-JD1 was first identified as a strongly lensed galaxy at a photometric redshift of 9.6 behind the MACS1149+2223 cluster \citep{Zheng+12} and later at 9.44$\pm$0.12 based on improved photometry (\citealt{Zheng+17}, see also \citealt{Hoag+18}). A confirmation of the redshift being above nine came from the detection of the [OIII]88$\mu$m emission line with the Atacama Large Millimeter Array (ALMA), $z$\,=\,9.1096\,$\pm$\,0.0006 \citep{Hashimoto+18}. The combined \textit{Hubble Space Telescope} (HST) and \textit{Spitzer} photometry is consistent with the spectral energy distribution (SED) of a young (<10 Myr old) plus a mature (290-512 \,Myr, \citealt{Hashimoto+18}, \citealt{Laporte+21a}) stellar population. The latter would explain the flux excess measured in the IRAC 4.5$\mu$m band relative to the 3.6$\mu$m band as being due to the rest-frame 0.4$\mu$m Balmer break, placing the epoch of formation of this galaxy at a redshift of 15 or above. New near-IR JWST imaging and spectroscopic observations have reported no evidence for the presence of a flux excess at 4.5 $\mu$m, excluding the existence of a prominent Balmer break produced by a dominant mature stellar population (\citealt{Bradac+23}, \citealt{Stiavelli+23}). The near-IR spectroscopy of JWST has confirmed that MACS1149-JD1 is a dust-free galaxy (H$\gamma$/H$\beta$\,=\,0.50$\pm$0.03) with a sub-solar metallicity (12\,+\,$\log(\mathrm{O/H}$)\,=\,7.88\,$\pm$\,0.05, \citealt{Stiavelli+23}). High-angular resolution [OIII]88$\mu$m observations have measured a velocity field consistent with a disk with a mass of 0.65$^{+1.37}_{-0.40}\times$10$^9$\,$M_{\odot}$ \citep{Tokuoka+22}. Finally, while MACS1149-JD1 is apparently bright due to its magnification, its intrinsic UV luminosity ($M_\mathrm{UV}$\,=\,$-$19.2 for a magnification of 11.5) places it in the intermediate luminosity range of the UV luminosity function at a redshift of nine \citep{Perez-Gonzalez+23b}.  MACS1149-JD1 is a good prototype to investigate the ionizing properties of similar sources in the early Universe.  

The Mid-Infrared Instrument (MIRI; \citealt{Rieke+15,Wright+15,Wright+23}) on board JWST is the only instrument that can provide spectroscopic observations of the H$\alpha$ emission line of galaxies at redshift above approximately seven \citep{Alvarez-Marquez+19_mrs} as well as rest-frame optical and near-IR imaging. This paper presents the first detection of the H$\alpha$ emission line and rest-frame optical ($\sim$0.55$\mu$m) imaging of the lensed MACS1149-JD1 galaxy at a redshift of 9.11. Section \ref{Sec:obs_cal} introduces the MIRI observations and calibrations. Section \ref{Sect:MIRI_phot_spec} presents the H$\alpha$ spectrum and the photometry and line flux measurements. Section \ref{Sec:results_dis} presents the results and discussion, which includes the stellar and ionized gas distribution (Sect. \ref{Sec:Stellar_ionized_dist}); the instantaneous SFR (Sect. \ref{Sec:SFR}); the ionizing photon production efficiency (Sect. \ref{Sec:ion_phothon_eff}); the equivalent width of H$\alpha$ (Sect. \ref{Sec:EW_Ha}) and its relation with the ionizing photon production efficiency (Sect. \ref{Sec:ew-photon}); and the kinematics of the ionized gas (Sect. \ref{Sect:kinematics}). Finally, Section \ref{Sec:conclusion_Summary} gives the summary and conclusions. Throughout this paper, we assume a Chabrier initial mass function (IMF; \citealt{Chabrier+03})  and a flat $\Lambda$CDM cosmology with $\Omega_\mathrm{m}$\,=\,0.310  and H$_0$\,=\,67.7\,km\,s$^{-1}$\,Mpc$^{-1}$ \citep{PlanckCollaboration18VI}. For this cosmology, 1 arcsec corresponds to 4.522\,kpc at $z$\,=\,9.1092, and the luminosity distance is $D_\mathrm{L}$\,=\,95331.2\,Mpc. For these cosmological parameters and redshift, the age of the Universe corresponds to 535\,Myr. A magnification factor of 11.5$^{+3.3}_{-3.1}$, mean of seven models \citep{Zheng+17}, was assumed when converting observed fluxes and sizes into intrinsic fluxes and sizes. We use vacuum emission line wavelengths throughout the paper.  

\section{MIRI observations and data calibration}\label{Sec:obs_cal}

MACS1149-JD1 was observed with MIRI on the 28 April and 12 May 2023 as part of the MIRI European Consortium guaranteed time observations (program ID\,1262). The observations are composed of MIRI imaging (MIRIM; \citealt{Bouchet+15}) and integral field spectroscopy using the Medium Resolution Spectrograph (MRS; \citealt{Wells+15,Argyriou+23}). The MIRIM observations were performed using the F560W filter covering the rest-frame optical at $\sim$\,0.55\,$\mu$m. It has a total on-source integration time of 2675 seconds using the FASTR1 readout mode. The observational setup consists of a four-point dither pattern and two integrations of 120 groups each. The MIRIM F560W observations were calibrated with version {1.12.3} of the JWST pipeline \citep{bushouse_pipeline} and context {1188} of the Calibration Reference Data System (CRDS). In addition to the general procedure, our data calibration process includes additional steps to correct for striping artifacts and background gradients (see \citealt{Alvarez-Marquez+23-SPT,Perez-Gonzalez+23a} for details). The final dithered F560W image has a pixel size of 0.06$\arcsec$.

\begin{figure*}
\centering
   \includegraphics[width=\linewidth]{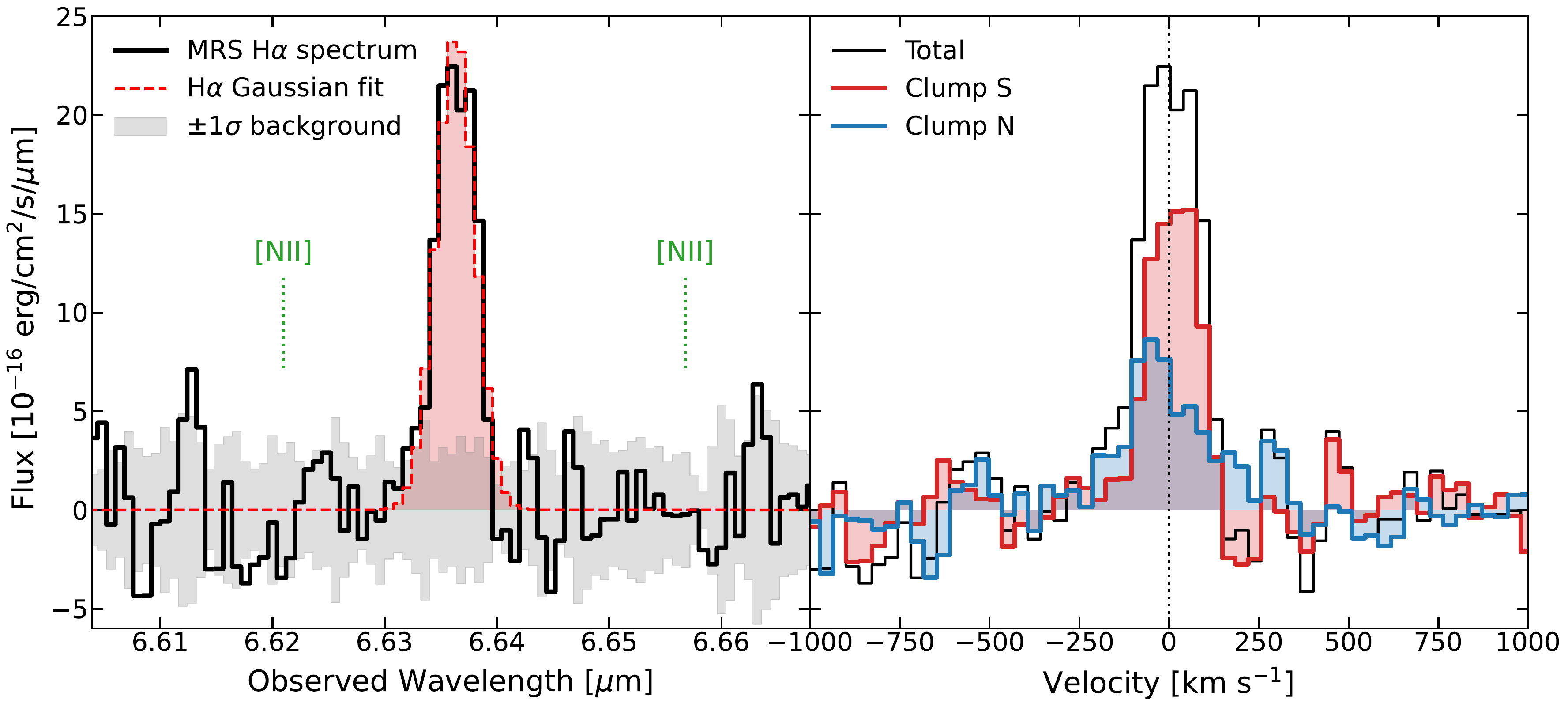}
      \caption{MRS 1LONG spectrum of MACS1149-JD1 centered on the H$\alpha$ emission line. The left panel shows the total integrated H$\alpha$ spectrum. The black line represents the integrated MRS spectrum of MACS1149-JD1 extracted using the total aperture (see Figure \ref{fig:HaLineMap}). The red dashed line and area are the best one-component Gaussian fit. The gray area indicates the $\pm$1$\sigma$ calculated as the standard deviation from nine different background spectra. The green dashed line shows the location of the [NII]6550,6565$\AA$ emission lines. The right panel shows the H$\alpha$ spectra in velocity space for the two spatially separated clumps identified in the H$\alpha$ line map. The black line represents the integrated MRS spectrum of MACS1149-JD1 extracted using the total aperture. The red line and area indicate the integrated MRS spectrum of clump S. The blue line and area show the integrated MRS spectrum of clump N.}
         \label{fig:HaLine_total}
\end{figure*}

The MRS observations were performed in the LONG band that simultaneously covers the wavelength ranges of 6.53$-$7.65\,$\mu$m, 10.02$-$11.70\,$\mu$m, 15.41$-$17.98\,$\mu$m and 24.19$-$27.9\,$\mu$m for channels 1, 2, 3 and 4, respectively. We only used the MRS Channel 1 for this work, which includes the spectral range of the redshifted H$\alpha$ emission line together with fainter emission lines such as [NII]6550,6565$\AA$ and [SII]6718,6733$\AA$. An inspection of the other MRS channels suggested no detection of additional emission lines. The on-source integration time is 22743 seconds, distributed in eight dither positions using the two available (Negative and Positive) four-point dither patterns. For each dither position, a total of six integrations were obtained using the SLOWR1 readout mode with 19 groups each. The MRS observations were processed with version 1.11.2 of the JWST calibration pipeline and context 1100 of the CRDS. In general, we followed the standard MRS pipeline procedure \citep{MRSpipeline} with additional customized steps to improve the quality of the final MRS-calibrated products (see \citealt{Alvarez-Marquez+23-SPT,Bosman+23} for details). In addition, we implemented a 1/f noise (correlated noise in the vertical direction of the detector) subtraction following \cite{Perna+23}, and we turned on the \texttt{outlier$\_$detection} step in stage three of the JWST pipeline. The final 3D spectral cube of channel 1LONG has a spatial and spectral sampling of 0.13"\,$\times$\,0.13"\,$\times$\,0.8\,nm \citep{Law+23} and a resolving power of $\sim$3500 ({85\,km\,s$^{-1}$}, \citealt{Labiano+21, Jones+23}).

Finally, we corrected the astrometry in the MIRIM F560W image and the MRS 1LONG 3D spectral cube. The F560W image was realigned by measuring the centroid of three field stars with available GAIA DR3 \citep{GaiaCollaboration+22} coordinates. To align the MRS cube, we used the two available GAIA DR3 stars in the field of view (FoV) of the MIRIM F770W and F1000W images taken simultaneously with the MRS observations. The final uncertainty in the data set alignment is less than a pixel in the MIRIM drizzle images (i.e., $<$\,60\,mas). Figure \ref{fig:HaLineMap} shows the MIRIM F560W images and the MRS H$\alpha$ line map together with an archival WFPC3/HST F160W image. The H$\alpha$ line map was generated by integrating the MRS 1LONG cube in the velocity range, -150\,<\,$v$\,[km\,s$^{-1}$]\,<\,150, and taking the H$\alpha$ emission line peak {derived by the Gaussian line fit} as a reference.

\section{MIRI photometric and spectroscopic measurements}\label{Sect:MIRI_phot_spec}

The MIRIM imaging and MRS spectroscopy detected the rest-frame optical ($\sim$\,0.55\,$\mu$m) and H$\alpha$ emission of MACS1149-JD1 (see Figure \ref{fig:HaLineMap}). The H$\alpha$ line map indicates the existence of two spatially resolved H$\alpha$ emitting regions (see Figure \ref{fig:HaLineMap}). These regions are spatially coincident with the morphology detected in the HST F160W image, which is composed of a bright clump in the south (henceforth, clump S) and a secondary fainter emission in the north (henceforth, clump N). This structure is compatible with the one recently published using the high-resolution near-IR NIRcam images \citep{Bradac+23,Stiavelli+23}. Three unresolved clumps (C1, C2, and C3) and an extended emission (G) have been identified in the NIRCam images \citep{Bradac+23}. In our lower angular resolution F560W image and H$\alpha$ line map, clump S is the combination of clumps C2 and C3 together with the extended G emission, and clump N is associated with C1. A detailed account of how the H$\alpha$ and F560W flux measurements were derived for the integrated galaxy and for the independent regions is given in the following sections.

\subsection{H$\alpha$ spectra and line fluxes}

We generated the 1D H$\alpha$ spectrum in different apertures to obtain the spectrum for MACS1149-JD1 and for each of the clumps. First, we extracted the integrated H$\alpha$ spectrum using an elliptical aperture with a semi-major axis of 0.5\arcs, a semi-minor axis of 0.325\arcs, a position angle of 115\degr, and a center at position with RA\,[deg]\,=\,177.389945 and DEC\,[deg]\,=\,+22.412722 (see the black dashed ellipse in Figure~\ref{fig:HaLineMap}). We also extracted nine 1D spectra using the same aperture at different positions of the MRS FoV clean of emission from the MACS1149-JD1 galaxy. We combined these spectra to generate the 1D median and standard deviation of the local background. {The median, which is compatible with zero, was subtracted from the H$\alpha$ spectrum with the goal of removing any systematic residual feature left in the MRS calibration process.} The H$\alpha$ spectrum was modeled by a one-component Gaussian function and a second-order polynomial to fit the emission line and any residual background gradient, respectively. The total H$\alpha$ spectrum is shown in Figure~\ref{fig:HaLine_total} together with the Gaussian fit and the 1$\sigma$ uncertainty. The observed H$\alpha$ line is centered at a wavelength of 6.6363\,$\pm$\,0.0002\,$\mu$m, corresponding to a redshift of 9.1092\,$\pm$\,0.0002 for MACS1149-JD1, in agreement with previous far-IR line observations (see Sect. \ref{Sect:kinematics}). The H$\alpha$ line was spectrally resolved and presented an intrinsic full width half maximum (FWHM) of 163\,$\pm$\,12\,km\,s$^{-1}$ after instrumental broadening correction \citep{Labiano+21}. The total observed H$\alpha$ flux is (1.05\,$\pm$\,0.07)\,$\times$\,10$^{-17}$\,erg\,s$^{-1}$\,cm$^{-2}$, and it was detected with a signal-to-noise ratio of 15. Fainter [NII]6550,6565$\AA$ and [SII]6718,6733$\AA$ emission lines were not detected with an observed 3$\sigma$ upper limit of 2.1\,$\times$\,10$^{-18}$\,erg\,s$^{-1}$\,cm$^{-2}$ (in agreement with the metallicity provided by \citealt{Stiavelli+23}). The H$\alpha$ fluxes were corrected for aperture losses, assuming that the two clumps identified in the H$\alpha$ line map are unresolved sources for the MRS angular resolution. The MRS point spread function (PSF) has an FWHM equal to 0.37\arcs{}\,$\times$\,0.31\arcs at 6.64$\mu$m \citep{Argyriou+23}. The percentage of flux outside the selected aperture is 39\% of the total using the latest MRS PSF models (Papatis et al. in prep.). The uncertainties on the derived emission line parameters, such as the line FWHM, flux, central wavelength, were estimated using a Monte Carlo simulation. The noise of the spectrum was measured as the root mean square of the continuum surrounding the emission line. This noise was used to generate new spectra (n\,=\,1000), where a random Gaussian noise with a sigma equal to the rms was added to the original spectrum before the lines were fitted again. The final uncertainty is the standard deviation of all the individual measurements. 

The 1D spectral extractions of the H$\alpha$ in clumps N and S used elliptical apertures of semi-major axis of 0.225\arcs and 0.275\arcs, a position angle of 130 and 30 deg, and a semi-minor axis of 0.175\arcs and 0.225\arcs. The clumps were respectively centered at the (RA\,[deg], DEC\,[deg]) positions (177.3899805, 22.41278811) and (177.3899325, 22.4126886). These elliptical apertures are represented with black dotted lines in Figure \ref{fig:HaLineMap}. The H$\alpha$ spectra and fluxes were corrected by aperture losses, and by the contamination of the companion clump taken into account the MRS PSF and that both clumps are unresolved. The H$\alpha$ line emissions derived for clump N and S are compared with the integrated H$\alpha$ line in Figure \ref{fig:HaLine_total}, and their fluxes and properties are presented in Table \ref{tab:PhotometryFluxes}. The total H$\alpha$ flux and the sum of the individual clump fluxes are in agreement within the uncertainties. This suggests that the H$\alpha$ emission is mainly dominated by the two unresolved clumps, excluding any diffuse component. 

\subsection{MIRIM F560W and HST F160W photometry}

The MIRIM F560W total photometry was generated using the same aperture as for the 1D spectral extraction of H$\alpha$. The local background level and standard deviation was estimated in a circular annulus centered at the same position as the total aperture and with an inner radius of 0.7\arcs and outer radius of 1.5\arcs. The measured flux and uncertainties were corrected by aperture losses, considering that the emission is unresolved and centered at the position of the bright emission. The percentage of flux outside the selected elliptical aperture is 37\% of the total, assuming the empirical MIRIM F560W PSF with a $\sim$0.2\arcs FWHM, as derived for the MIRI Deep Imaging Survey (MIDIS; \citealt{Boogaard+23}). We also scaled the photometric uncertainties in order to take into account the correlated noise introduced by drizzling in the MIRIM F560W image. We used 1000 apertures on blank regions to measure the rms on both the nominal and drizzled images. This effect was estimated to be a factor of {2.192} that needs to be applied to the errors calculated in the drizzled F560W image (\"Ostlin et al. in prep.). The final observed MIRIM F560W flux of MACS1149-JD1 is {0.34\,$\pm$\,0.03\,$\mu$Jy}. Following the same methodology, we obtained an HST F160W total flux of 0.224\,$\pm$\,0.002\,$\mu$Jy. To correct for aperture losses, we used a PSF generated using three stars of the HST image FoV. The percentage of flux outside the selected elliptical aperture is 28\%. The HST F160W photometry is in perfect agreement with the one obtained from the NIRCam and NIRISS imaging at 1.5$\mu$m (0.230\,$\pm$\,0.017\,$\mu$Jy \citealt{Bradac+23}).

The HST F160W and MIRIM F560W fluxes for clump N and S were calculated by applying a methodology similar to the one used in the 1D extraction of the H$\alpha$ spectrum. The photometry (see Table \ref{tab:PhotometryFluxes}) was obtained for clumps N and S after decontamination of the contribution of the companion clump in the aperture and by applying the corresponding aperture corrections for unresolved sources. In addition, we modeled the surface brightness distribution in F560W using GALFIT \citep{Peng+02}. Following the structure found in NIRCam \citep{Bradac+23,Stiavelli+23}, we used two unresolved sources to fit the central emission of the galaxy (corresponding to C2 and C3) and a S\'ersic profile to fit the diffuse emission (corresponding to G). Clump C1, which is mainly detected as an unresolved source in the rest-frame UV NIRCam images, is not detected above the residuals of the combined emission of the C2 and C3 clumps and the diffuse component. The total flux given by the sum of all individual components is {0.33\,$\pm$\,0.03\,$\mu$Jy}, which is in perfect agreement with the one obtained in the total aperture. The flux for clump S, which is the combination of the unresolved components after correcting for aperture plus the additional contribution within the aperture of the diffuse emission, is {0.26\,$\pm$\,0.02\,$\mu$Jy} following GALFIT analysis. These values are in agreement within the uncertainties with the ones using the general methodology and are given in Table \ref{tab:PhotometryFluxes}. We did not implement aperture correction for the F560W flux of clump N in the general methodology because the GALFIT analysis showed that the aperture of clump N is dominated by diffuse emission. The GALFIT model for the HST F160W image does not converge to a solution, so we assumed that the UV light is dominated by the bright clumps identified in the NIRCam images (e.g., \citealt{Stiavelli+23}), and we implemented aperture corrections in both clumps.      

\begin{table*}[!ht]
\caption{Photometry and H$\alpha$ fluxes and main physical properties for MACS1149-JD1 and each of its clumps.}
\centering
\begin{tabular}{lccc}
\hline
 & Total & Clump S & Clump N \\
\hline
Flux HST F160W [$\mu$Jy] & 0.224$\pm$0.002 & 0.162$\pm$0.001 & 0.042$\pm$0.001 \\
Flux MIRIM F560W [$\mu$Jy] & {0.34$\pm$0.03} & {0.25$\pm$0.02} & {0.032$\pm$0.007} \\
Flux H$\alpha$ [$\times$10$^{-18}$\,erg\,s$^{-1}$\,cm$^{-2}$] & 10.5$\pm$0.7 & 6.3$\pm$0.4 & 4.6$\pm$0.9 \\
H$\alpha$ peak [$\mu$m] & 6.6363$\pm$0.0002 & 6.6367$\pm$0.0001 & 6.6359$\pm$0.0005 \\
FWHM H$\alpha$ [km\,s$^{-1}$]& 163$\pm$12 & 131$\pm$9 & 267$\pm$77 \\
$M_{\mathrm{UV}}$ [AB\,mag] & $-$19.2 & $-$18.9 & $-$17.4 \\
SFR$_\mathrm{H_{\alpha}}$\,[$M_{\odot}$\,yr$^{-1}$]$^{(1)}$ & 3.2$\pm$\,0.3 & 1.9\,$\pm$\,0.2 & 1.4$\pm$0.3 \\
$\log(\zeta_\mathrm{ion}$ [$\mathrm{Hz\,erg^{-1}}$]) & 25.55$\pm$0.03 & 25.47$\pm$0.03 & 25.91$\pm$0.09 \\
EW$_{0}$\,(H$\alpha$) [$\AA$] & {726$^{+660}_{-182}$} & {531$^{+300}_{-96}$} & {>1951} \\
\hline
\label{tab:PhotometryFluxes}
\end{tabular}
\tablefoot{The table presents observed fluxes that have not been corrected by magnification. The other parameters are not affected by the magnification, except for the cases of $M_{\mathrm{UV}}$ and SFR$_\mathrm{H_{\alpha}}$, which use a magnification correction of 11.5. The $^{(1)}$ SFR$_\mathrm{H_{\alpha}}$ were calculated for a metallicity of 0.28\,Z$_{\odot}$ (see \citealt{Reddy+18} and Sect. \ref{Sec:SFR}).}
\end{table*}

\section{Results and discussion}\label{Sec:results_dis}

We detected and measured the spatially resolved H$\alpha$ and rest-frame optical emission in a galaxy at redshift above nine, that is, during the initial stages of the EoR. Combined with rest-frame UV ancillary data, the new MIRI data provide direct constraints for important properties of MACS1149-JD1, such as the spatially resolved instantaneous SFR, the ionizing photon production efficiency, the H$\alpha$ equivalent width, the ionized gas kinematics, and dynamical mass estimates. MACS1149-JD1 has an intrinsic UV luminosity ($M_\mathrm{UV}$\,=\,$-$19.2 for a magnification of 11.5)  in the intermediate luminosity range of the UV luminosity function at a redshift of nine \citep{Perez-Gonzalez+23b}, and it can therefore be considered a good prototype to investigate the ionizing properties of similar sources at this early epoch of the Universe. The main results of this work are presented in the following sections.

\subsection{Stellar and ionized gas distribution}\label{Sec:Stellar_ionized_dist}

Figure~\ref{fig:HaLineMap} shows the HST F160W and  MIRIM F560W images and the MRS H$\alpha$ line map of MACS1149-JD1 at $z$\,=\,9.11. The F160W image traces the rest-frame UV emission at $\sim$0.16\,$\mu$m, which is mainly dominated by the young stellar population. Recently, high-resolution NIRCam observations decomposed MACS1149-JD1 into three unresolved clumps and a diffuse emission at rest-frame UV \citep{Bradac+23,Stiavelli+23}. This structure is compatible with the one seen in HST, which shows a bright clump in the south, clump S, and a secondary fainter emission in the north, clump N. 

The MIRIM F560W image traces the rest-frame optical ($\sim$\,0.55\,$\mu$m), combining the stellar continuum and emission lines. The strong optical emission lines, such as [OIII]4960,5008$\AA$, are located close to the edge of the filter bandpass, where the transmission decreases rapidly. If we consider the derived H$\alpha$ (see Sect.~\ref{Sect:MIRI_phot_spec}) and assume the low-metallicity ($\sim$0.2\,$Z_{\odot}$) templates from \cite{Alvarez-Marquez+19_mrs}, the contributions of the [OIII]4960,5008$\AA$ emission lines would be {about 35\%} of the total flux measured in the F560W image. Their contribution drop to {about 15\% and 7\%} for the assumption of metal-poor galaxies with metallicities 0.04 and 0.02\,$Z_{\odot}$. In the case of having an extreme [OIII]5008$\AA$/H$\alpha$ flux ratio of 3.5 \citep{Stiavelli+23}, the [OIII]4960,5008$\AA$ contribution would be up to {about 60\%}. The flux in the F560W image is dominated by the emission of clump S, which could be decomposed into two unresolved sources and a diffuse emission (see Sect. \ref{Sect:MIRI_phot_spec}), while clump N is dominated by the outskirt regions of the diffuse emission.

The ionized gas traced by the H$\alpha$ emission shows a spatial structure similar to that of the rest-frame UV. The peak emission coincides with that of the HST F160W and MIRIM F560W images, clump S, while an extension toward the north is detected, clump N, which is coincident with the emission of a rest-frame UV clump \citep{Bradac+23,Stiavelli+23}. Moreover, the ALMA [OIII]88$\mu$m emission line observation presents an elongated structure of the size 0.82\arcsec${}\,\times$\,0.3\arcs, in agreement with the extension of the H$\alpha$ emission. This indicates that the stellar population and the nebular emission are located in the same regions in MACS1149-JD1, where H$\alpha$ traces the presence of young stellar populations. 

\subsection{Instantaneous star-formation rate}\label{Sec:SFR}

Considering no significant internal extinction is present, we derived the instantaneous SFR directly from the H$\alpha$ flux following \citep{Kennicutt-Evans+12}:

\begin{equation}
    \mathrm{SFR}(M_{\odot}\,\mathrm{yr}^{-1}) = 5.37 \times 10^{-42} \times L(\mathrm{H\alpha, erg\, s^{-1}}) \times (1 - f_\mathrm{esc})^{-1}
\end{equation}
\noindent
for a Chabrier IMF and a given escape fraction ($f_\mathrm{esc}$) for the ionizing photons. {As we are not in the position to derive $f_\mathrm{esc}$ for this source, we assumed $f_\mathrm{esc}$\,=\,0 for comparison with other works.} Under this assumption, the SFR corresponds to 5.3\,$\pm$\,0.4\,$M_{\odot}$\,yr$^{-1}$. For lower than solar metallicities, a slightly lower H$\alpha$ luminosity to SFR conversion factor (3.236\,$\times$\,10$^{-42}$ for 0.28\,$Z_{\odot}$, \citealt{Reddy+18}) should be used. Applying this factor, the SFR  is 3.2\,$\pm$\,0.2\,$M_{\odot}$\,yr$^{-1}$ (see Table \ref{tab:PhotometryFluxes} for the SFRs of clumps N and S). These SFR values agree with those derived from the UV continuum using the HST F160W image (i.e., 0.158\,$\mu$m rest-frame). The conversion of the UV flux to the SFR depends not only on the metallicity but also on the star-formation history (SFH), with normalization factors that have a large dependence on the length of the star-formation process (e.g., \citealt{Calzetti+13}). For MACS1149-JD1, with an integrated F160W observed flux of 0.224\,$\pm$\,0.002\,$\mu$Jy, an SFR of 6.3, 2.7, and 1.9\,$M_{\odot}$\,yr$^{-1}$ was derived for solar metallicity and a constant SFH of 2\,Myr, 10\,Myr, and more than 100\,Myr, respectively. Finally, the H$\alpha$-based SFR agrees with the far-IR estimate (SFR([CII]158$\mu$m)= 5.7\,$M_{\odot}$\,yr$^{-1}$) based on the detection (4.6$\sigma$) of the [CII]158$\mu$m emission line \citep{Carniani+20} and assuming that the SFR to [CII]158$\mu$m luminosity relation for low-z galaxies \citep{DeLooze+14} is also valid for high-z galaxies \citep{Schaerer+20}. 

The good agreement between the H$\alpha$-, UV-, and [CII]158$\mu$m-based SFRs already provides relevant information about the system. While the SFR derived from the H$\alpha$ line {traces the presence of massive young ionizing stars} (i.e., less than 10 Myr old), the UV flux can also trace the non-ionizing continuum of less massive stars and therefore potentially longer periods of time in the recent SFH of the galaxy. The first qualitative conclusion is that internal extinction is very low and is not playing a relevant role. This is supported by recent NIRSpec/JWST measurement of the H$\beta$/H$\gamma$ ratio that is in agreement with Case B recombination \citep{Stiavelli+23}. Otherwise, the H$\alpha$-derived SFR would be much larger than the UV-based SFR due to the increasing extinction effect toward shorter wavelengths. On the other hand, the H$\alpha$-to-UV SFR ratio is close to one for ages younger than 10 Myr only. Thus, a young, unobscured stellar population appears to dominate the UV spectrum and the ionization of the ISM in this galaxy. {The mass of the young stellar population can also be estimated from the prediction of the number of ionizing photons produced by the young stars. The integrated H$\alpha$ luminosity provides a total number of ionizing photons equal to $N_\mathrm{LyC}$\,=\,7.4\,$\times$\,10$^{53}$\,ph\,s$^{-1}$ (see Sect.~\ref{Sec:ion_phothon_eff}). This ionizing radiation can only be produced by a young stellar burst with a stellar mass equal to $\sim$10$^7$\,$M_{\odot}$} (e.g., \citealt{Stanway+Eldridge-23}). As the UV emission is distributed in several compact, bright clumps \citep{Bradac+23}, the stellar mass in each of these clumps is predicted to be of the order of a few $\times$\,10$^6$\,$M_{\odot}$.

\subsection{Ionizing photon production efficiency}\label{Sec:ion_phothon_eff}

\begin{figure}
\centering
   \includegraphics[width=\linewidth]{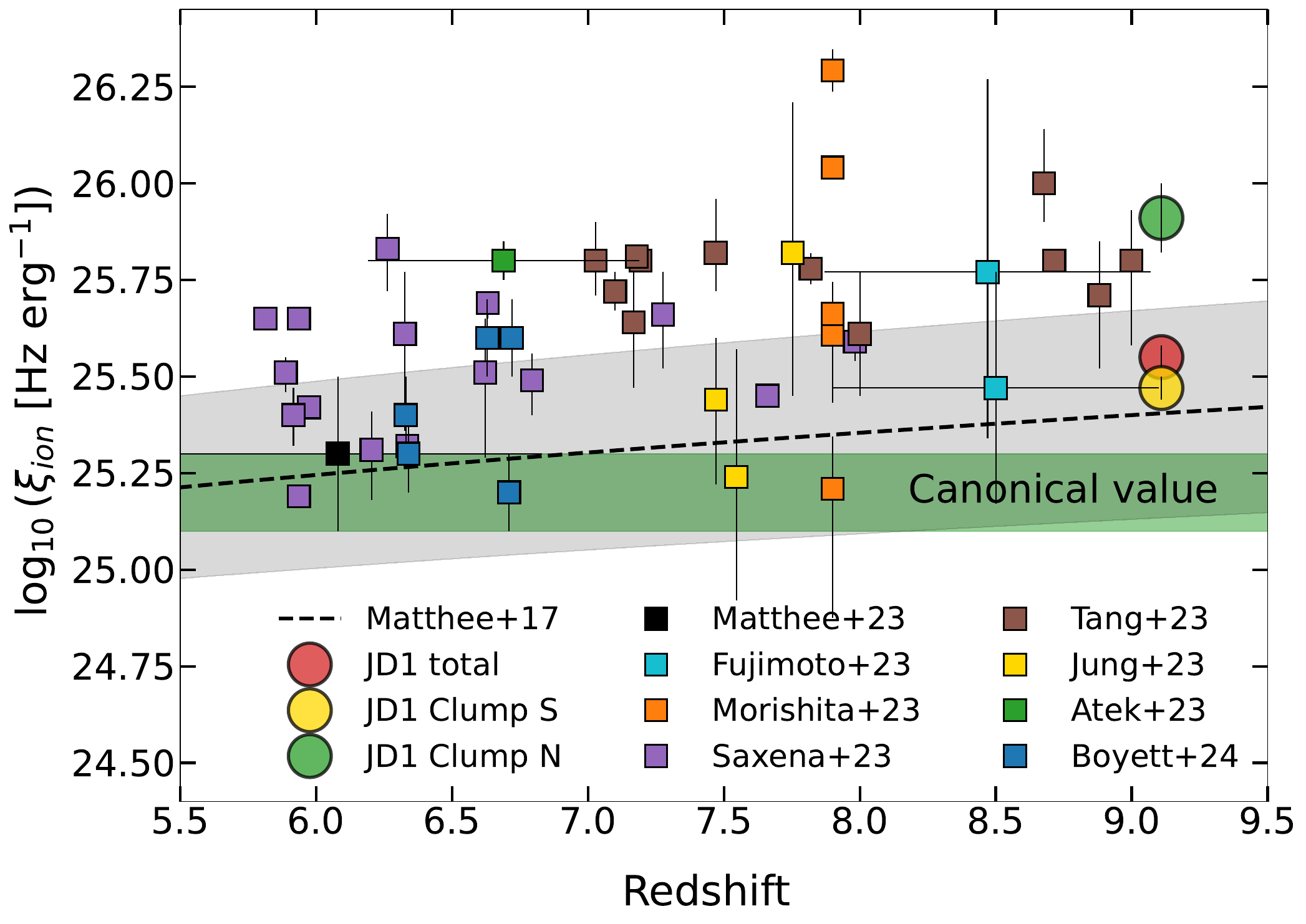}
      \caption{Ionizing photon production efficiency as a function of redshift. MACS1149-JD1 is represented by circles, where we distinguish between the values of the integrated galaxy (red) and the spatially resolved clumps N (green) and S (yellow). The squares indicate galaxies spectroscopically identified at $z$\,$\gtrsim$\,6 with JWST \citep{Fujimoto+23,Jung+23,Morishita+23,Saxena+23,Matthee+23,Tang+23,Atek+23,Boyett+24}. The green area shows the canonical value for the ionizing photon efficiency \citep{Robertson+13}. The black line and gray area show the variation of the photon efficiency with redshift and its uncertainty \citep{Matthee+17}.}
         \label{fig:photon-z}
\end{figure}

The ionizing photon production efficiency is given as the ratio of the ionizing to the non-ionizing UV flux. For an H$\alpha$ luminosity and assuming recombination, the ratio is given as 

\begin{equation}
\zeta_\mathrm{ion}(\mathrm{Hz\,erg^{-1}})= \frac{N_\mathrm{LyC} (\mathrm{ph\,s^{-1}})}{L_\mathrm{UV}(\mathrm{\,erg\,s^{-1}\,Hz^{-1}})}
\end{equation}

\noindent
with 

\begin{equation}\label{eq:4}
N_\mathrm{LyC}(\mathrm{ph\,s^{-1}})= 7.5 \times 10^{11} \times \frac{L(\mathrm{H\alpha}, \mathrm{erg\,s^{-1})}}{(1-f_\mathrm{esc})},
\end{equation}

\noindent
where $f_\mathrm{esc}$ are the fractions of ionizing photons  escaping into the IGM. The number of ionizing photons ($N_\mathrm{LyC}$) is given for an ISM with electron temperatures of 1.5\,$\times$\,10$^4$\,K and low densities (e.g., \citealt{Colina+91}). The ratio of H$\alpha$ luminosity to the number of ionizing photons has a slight dependence on the electron temperature. We chose the factor for a temperature of 1.5\,$\times$\,10$^4$\,K, measured in low-metallicity low-$z$ galaxies (see, e.g., \citealt{Alvarez-Marquez+19_mrs} and references therein). Assuming an escaping fraction of zero, the $\log(\zeta_\mathrm{ion}$) values are 25.55\,$\pm$\,0.03; 25.47\,$\pm$\,0.03; and 25.91\,$\pm$\,0.09 Hz\,erg$^{-1}$ for MACS1149-JD1 and for the spatially resolved clumps S and N, respectively (see Table~\ref{tab:PhotometryFluxes}). These values (see Figure~\ref{fig:photon-z}) are significantly larger than the canonical value (log($\zeta_\mathrm{ion}$)\,=\,25.2\,$\pm$\,0.1\,Hz\,erg$^{-1}$, \citealt{Robertson+13}) and larger than the values measured in bright ($M_\mathrm{UV}$\,$<$\,$-20$) and massive ($\log(M_\mathrm{star}/M_{\odot}$)\,$>$\,9.5) intermediate redshift (2$<$\,$z$\,$<$\,5) galaxies from the VANDELS sample \citep{Castellano+23}. Only VANDELS galaxies with extremely high, specific SFRs ($\log$(sSFR\,/\,yr$^{-1}$)\,$\sim$\,$-$7.5) show efficiencies  (25.5 Hz\, erg$^{-1}$) similar to that measured in MACS1149-JD1. The ionizing photon production in MACS1149-JD1 also lies slightly above the $z$\,=\,9.11 extrapolated value (log($\zeta_\mathrm{ion}$)\,=\,25.4\,Hz\,erg$^{-1}$) inferred as a function of redshift from galaxies at a redshift of 2.2 and the (1+$z$)$^{1.3}$ variation (\citealt{Matthee+17}, see Fig. \ref{fig:photon-z}). Also, lensed dwarfs at a redshift around two show lower photon efficiencies ($\sim$\,0.4 in absolute log units, \citealt{Emami+20}) independent of their stellar mass (7.8\,$<$\,log($M_\mathrm{star}/M_{\odot})$\,$<$\,9.8) or UV luminosity ($-$22\,$<$\,$M_\mathrm{UV}$\,$<$\,$-$17.3).

A comparison of the photon production efficiency of MACS1149-JD1 and its clumps with galaxies at higher redshifts ($z$\,$\gtrsim$\,6) is presented in Figure \ref{fig:photon-z}. Recent JWST programs have been able to measure the hydrogen Balmer lines (H$\beta$ or H$\alpha$) in galaxies at redshifts above five and therefore obtained a direct value for the photon production efficiency in galaxies during the late phases of the EoR. The photon efficiency for the EIGER galaxies with $M_\mathrm{UV}$\,=\,$-19.5$\,$\pm$\,0.1 at redshifts 6.25\,$<$\,\,$z$\,$<$\,6.93 corresponds to 25.31$^{+0.29}_{-0.16}$ \citep{Matthee+23}. Although the galaxies have a large scatter, this value is close to the canonical efficiency and on average is well below ($\times$1.4\,$-$\,4) that for MACS1149-JD1 and its two clumps. However, JADES galaxies covering the redshift range 5.8 to 8 and spanning the $-$17.0\,$\leq$\,$M_\mathrm{UV}$\,$\leq$\,$-$20.6 magnitude range have an average efficiency of about 25.5, \citep{Saxena+23,Boyett+24}, in agreement with the efficiency measured for the MACS1149-JD1 galaxy and clump S. We note that the JADES galaxies have an average UV magnitude $M_\mathrm{UV}$\,=\,$-$19.0, which is  similar to that of MACS1149-JD1 ($M_\mathrm{UV}$\,=\,$-$19.2). Additional measurements with JWST of galaxies at redshifts seven to nine showed photon production efficiencies similar to those measured in MACS1149-JD1 independent of their UV luminosity. The five galaxies behind the A2744 cluster at a redshift of 7.88 and with UV magnitudes $-$20.13\,$<$\,$M_\mathrm{UV}$\,$<$\,$-$19.28 have values spanning the 25.21\,$\leq$\,log($\zeta_\mathrm{ion})$\,$\leq$\,26.29\,Hz\,erg$^{-1}$ range, with an average of 25.91\,Hz\,erg$^{-1}$ \citep{Morishita+23}. The UV luminous ($-$22.2\,$<$\,$M_\mathrm{UV}$\,$<$\,$-$20.0) MIDIS H$\alpha$ emitters at redshifts seven to eight have an average log($\zeta_\mathrm{ion}$)\,=\,25.59$^{+0.06}_{-0.04}$\, Hz\,erg$^{-1}$ (Rinaldi et al. in prep.). The sample of CEERS NIRCam-selected spectroscopically confirmed galaxies at redshifts 7.8 to 9 \citep{Fujimoto+23} have median values of $\log(\zeta_{\mathrm{ion}})$\,=\,25.77$^{+0.50}_{-0.43}$\,Hz\,erg$^{-1}$. Finally, the faint galaxies ($M_\mathrm{UV}$\,>\,$-16.5$) present values of log($\zeta_{\mathrm{ion}}$)\,=\,25.8\,$\pm$\,0.05\,Hz\,erg$^{-1}$ higher than the MACS1149-JD galaxy but are in close agreement with clump N, which has a similar M$_{\mathrm{UV}}$ ($-$17.4). 

In summary, the ionizing photon efficiency in MACS1149-JD1 is similar to the value measured in other galaxies at similar redshifts and well above the canonical value and the range of values derived for the different populations of intermediate redshift galaxies. Recent results of the Sunrise Arc confirms the presence of young stellar clusters with a large photon production efficiency ($\log(\zeta_{\mathrm{ion}})$\,=\,25.7\,Hz\,erg$^{-1}$; \citealt{Vanzella+23,Adamo+24}). This could indicate that the clumps of MACS1149-JD1 host young stellar clusters because they present similar photon production efficiency.

\subsection{EW$_{0}$\,(H$\alpha$) and redshift evolution}\label{Sec:EW_Ha}

\begin{figure}
\centering
   \includegraphics[width=\linewidth]{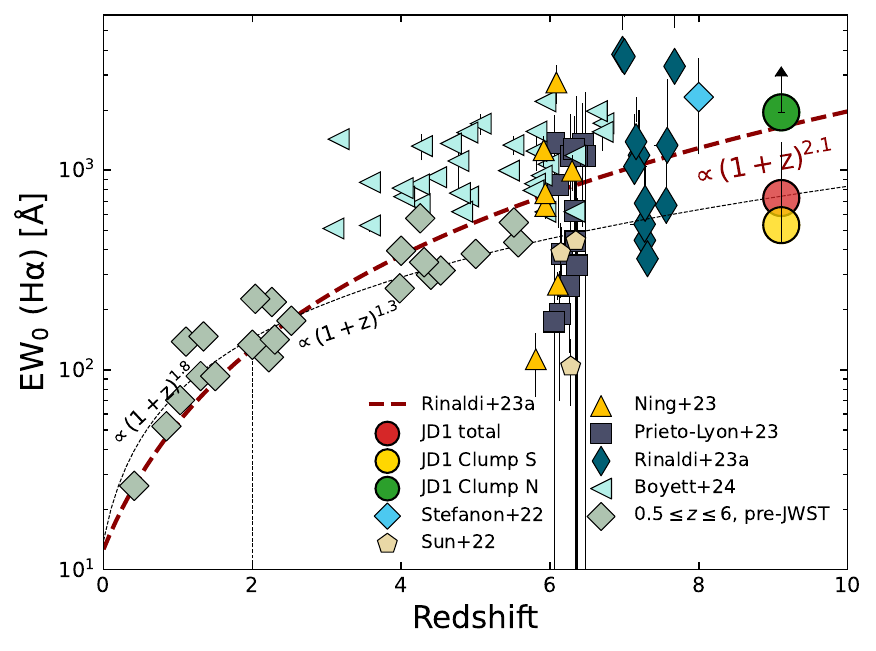}
      \caption{Evolution of the rest-frame equivalent width of H$\alpha$ as a function of redshift. MACS1149-JD1 is represented by circles, and we distinguish between the values of the integrated galaxy (red) and the spatially resolved clumps N (green) and S (yellow). {Additional data includes samples detected with JWST at redshifts of $3<z<7$ \citep{Boyett+24}; 6 \citep{Sun+22,Ning+23,Prieto-Lyon+23}; and 7$-$8 \citep{Rinaldi+23} as well as pre-JWST galaxies at redshifts 0.5 to 8}. The pre-JWST value at redshift eight corresponds to the median stacking of a sample of 102 Lyman-break galaxies \citep{Stefanon+22}. The dashed dark red line represents the best fit to all data points, including the MIDIS sources, and described by a single law EW$_{0}$\,(H$\alpha$)\,$\propto$\,(1+$z$)$^{2.1}$ by \cite{Rinaldi+23}. The dotted broken line represents previous fits with pre-JWST data up to a redshift of six \citep{Faisst+16}.
      }
         \label{fig:EW}
\end{figure}

We estimated the rest-frame equivalent width of H$\alpha$ from the F560W and H$\alpha$ fluxes. First, we measured the rest-frame optical ($\sim$0.55\,$\mu$m) continuum by subtracting the contribution of [OIII]4960,5008$\AA$ emission lines from the F560W flux. We considered a [OIII]5008$\AA$/H$\alpha$ flux ratio equal to two, in agreement with the observed [OIII]5008$\AA$/H$\beta$ flux ratios of high-z ($z$\,>\,5) galaxies with JWST \citep{Matthee+23,Cameron+23} and also similar to low-metalicity ($\sim$0.2\,Z$_{\odot}$) and low-redshift galaxies (\citealt{Alvarez-Marquez+19_mrs} and reference therein). We used a range in the [OIII]5008$\AA$/H$\alpha$ flux ratio between 1 to 3.5 to determine the uncertainties. The lowest value corresponds to the typical value measured in low-redshift metal-poor galaxies ($\sim$0.04\,Z$_{\odot}$, \citealt{Alvarez-Marquez+19_mrs} and references therein), while the highest value corresponds to the line ratio reported by \cite{Stiavelli+23} for clump S in MACS1149-JD1 considering no dust emission and Case B recombination. We used the high [OIII]5008$\AA$/H$\alpha$ flux ratio of 3.5 as an upper limit for our uncertainties. {The [OIII]5008$\AA$/H$\alpha$ flux ratios gave an OIII]4960,5008$\AA$ flux contribution of 17 (16)\%, 38 (32)\%, and 67 (56)\% to the measured F560W flux for the total 
(clump S) apertures, respectively.} Second, we assumed a flat optical continuum spectrum in $F_\mathrm{\nu}$ in order to extrapolate the derived optical ($\sim$0.55\,$\mu$m) continuum fluxes to the H$\alpha$ wavelength. Finally, the EW$_{0}$\,(H$\alpha$) was derived by the ratio between the H$\alpha$ flux and the continuum flux under the H$\alpha$. Table~\ref{tab:PhotometryFluxes} shows the EW$_{0}$\,(H$\alpha$) for MACS1149-JD1 and its clumps.        

The rest-frame equivalent width of the H$\alpha$ emission line for the MACS1149-JD1 is {726$^{+660}_{-182}$\,$\AA$. This equivalent width is lower} than the predicted extrapolation of the $(1+z)^{2.1}$ relation derived by the combination of pre-JWST measurements and the results of the JADES and MIDIS JWST surveys (see Figure~\ref{fig:EW}, \citealt{Rinaldi+23}). The EW$_{0}$\,(H$\alpha$) of MACS1149-JD1 is a factor of {two} lower than the extrapolated value at a redshift of 9.11, and it is in the range of the lower values measured in galaxies at redshifts six to eight recently identified with JWST (Figure \ref{fig:EW},  \citealt{Sun+22,Prieto-Lyon+23,Ning+23,Rinaldi+23}). {MACS1149-JD1 EW$_{0}$\,(H$\alpha$) is in agreement within the errors with the extrapolation of the $(1+z)^{1.3}$ redshift evolution derived from pre-JWST measurements with galaxies up to a redshift of six}. Lacking other measurements at redshifts above nine, MACS1149-JD1 could represent the lower end of a wide EW$_{0}$\,(H$\alpha$) distribution at redshifts 9$-$10, very much like the one already observed at redshifts between six and eight. 

The spatially resolved F560W imaging and H$\alpha$ spectroscopy identified two clumps with very different EW$_{0}$\,(H$\alpha$), covering the extremes of the range of values measured in high-z galaxies. Clump S has an EW$_{0}$\,(H$\alpha$) of {531$^{+300}_{-96}\AA$}, while clump N has a lower limit of {1951\,$\AA$}. We interprete it as the presence of different stellar populations in the two clumps. The emission in clump S is a combination of young stars ($<$\,10\,Myr) with a non-ionizing, more mature stellar population (see Sect. \ref{Sec:ew-photon}). On the other side, clump N has to be associated with a very young stellar {population}. Only {stars} with ages lower than {5\,Myr} can produce such a high EW$_{0}$\,(H$\alpha$). In addition, the integrated EW$_{0}$\,(H$\alpha$) for MACS1149-JD1 is close to that of clump S, indicating that the older, more mature stellar population must be concentrated in this clump, as already seen in the F560W image.

\subsection{log($\zeta_\mathrm{ion}$)\,$-$\,EW$_{0}$\,(H$\alpha$) relation and stellar populations}\label{Sec:ew-photon}

\begin{figure}
\centering
   \includegraphics[width=\linewidth]{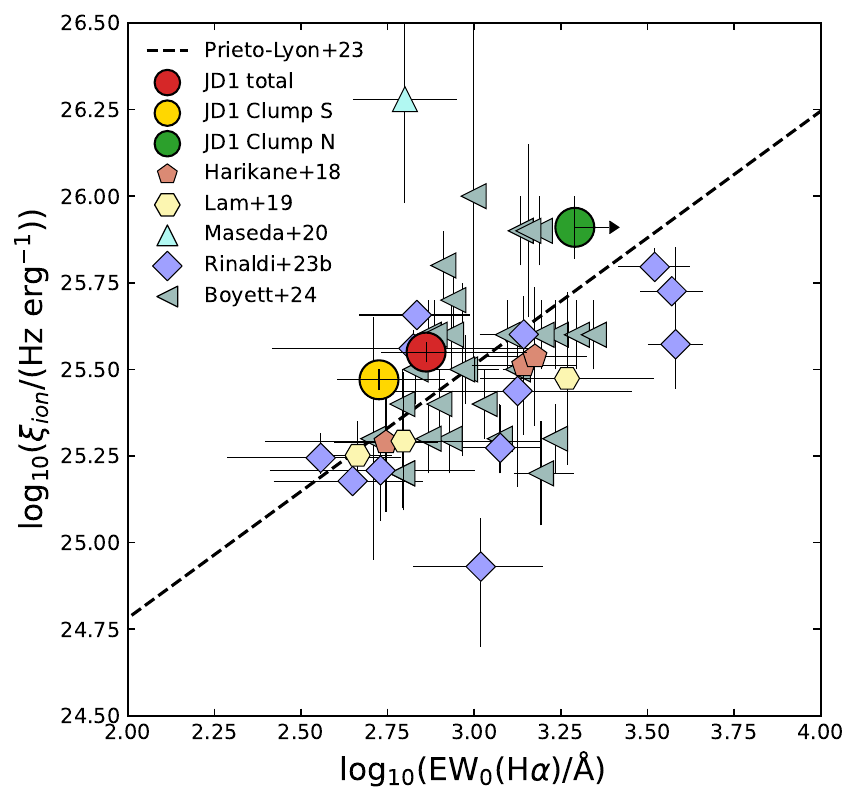}
      \caption{Ionizing photon production efficiency as a function of the rest-frame H$\alpha$ equivalent width. MACS1149-JD1 is represented by circles, and we distinguish between the values of the integrated galaxy (red) and the spatially resolved clumps N (green) and S (yellow). MACS1149-JD1 is compared with other samples of galaxies, including MIDIS H$\alpha$ emitters at $z$\,$\sim$\,7$-$8 \citep{Rinaldi+23}; faint Ly$\alpha$ emitters (LAEs) at $z$\,$\sim$\,4$-$6 \citep{Hashimoto+18, Lam+19, Maseda+20}{; and extreme emission line galaxies at $3<z<7$ \citep{Boyett+24}}. The dashed line represents the log($\zeta_\mathrm{ion}$)\,$-$\,EW$_{0}$\,(H$\alpha$) relation for galaxies at redshifts three to seven from the JWST GLASS and UNCOVER surveys \citep{Prieto-Lyon+23}.}
         \label{fig:EWHa-Xion}
\end{figure}

The ionizing photon production efficiency ($\zeta_\mathrm{ion}$) for MACS1149-JD1 and its clumps have high values for their EW$_{0}$\,(H$\alpha$) that are above the expected values derived from the relationship \citep{Prieto-Lyon+23} obtained for galaxies at redshifts three to seven (see Figure~\ref{fig:EWHa-Xion}). These values are also above those measured in the MIDIS H$\alpha$ emitters with similar EW$_{0}$\,(H$\alpha$). Thanks to the spatially resolved H$\alpha$ imaging, the location of the clumps and the integrated MACS1149-JD1 value in the log($\zeta_\mathrm{ion}$)\,$-$\,EW$_{0}$\,(H$\alpha$) plane can be understood as being the result of the spatial distribution of different stellar populations and of their contribution to the overall flux at these wavelengths. On one side, the values for the photon production above 25.5 Hz\,erg$^{-1}$ (in log($\zeta_\mathrm{ion}$) units) can only be obtained with {massive stellar bursts (10$^{6-7}$\,$M_{\odot}$)}. For less {massive stellar bursts}, the ionizing photon efficiency will be lower ($\leq$\,25.2\,Hz\,erg$^{-1}$), even for low-metallicity stellar populations, due to the stochastic sampling of the stellar initial mass function \citep{Stanway+Eldridge-23}. Thus, the high $\zeta_\mathrm{ion}$ values measured in the clumps already indicate the presence of {massive stellar bursts} in these regions of the galaxy. In addition, only instantaneous or constant star formation over short periods of time ($\sim$\,10\,Myr) are able to produce the high-ionizing photon production efficiency measured in MACS1149-JD1, even if the presence of binaries is invoked (\citealt{Eldridge+17},\citealt{Eldridge+Stanway+20}, \citealt{Stanway+20}). Thus, the intrinsic high value for the photon production efficiency already indicates the presence of {young massive stellar bursts} in different regions of MACS1149-JD1 and, consequently, in the galaxy overall as well. However, clumps S and N have a very different EW$_{0}$\,(H$\alpha$), with a low value ({531$^{+300}_{-96}\AA$}) measured for clump S and a very high value (lower limit of {1951\,$\AA$}) for clump N (see Table \ref{tab:PhotometryFluxes}). Moreover, as the continuum light distribution in MACS1149-JD1 is dominated by clump S, the integrated H$\alpha$ emission in MACS1149-JD1 also has a relatively low value ({726$^{+660}_{-182}$\,$\AA$}). The H$\alpha$ equivalent widths of about 2000\,$\AA$ can only be produced by very young instantaneous or constant bursts with ages of less than {5 Myr} \citep{Eldridge+17}, {while lower values around 500$\AA$ indicate the presence of a relatively older population of about 50\,Myr if a constant star formation is considered}. Thus, the presence of both {young massive stellar bursts} together with older stellar populations needs to be invoked to explain the position of MACS1149-JD1 and its clumps in the log($\zeta_\mathrm{ion}$)\,$-$\,EW$_{0}$\,(H$\alpha$) plane. Similar conclusions have been obtained from recent NIRCam imaging \citep{Bradac+23}. Additional combinations of imaging covering longer wavelengths and emission line diagnostics will allow for further constraining of the stellar populations and their distribution in this galaxy.

\subsection{Kinematics of the ionized gas}\label{Sect:kinematics} 

\begin{figure}
\centering
   \includegraphics[width=\linewidth]{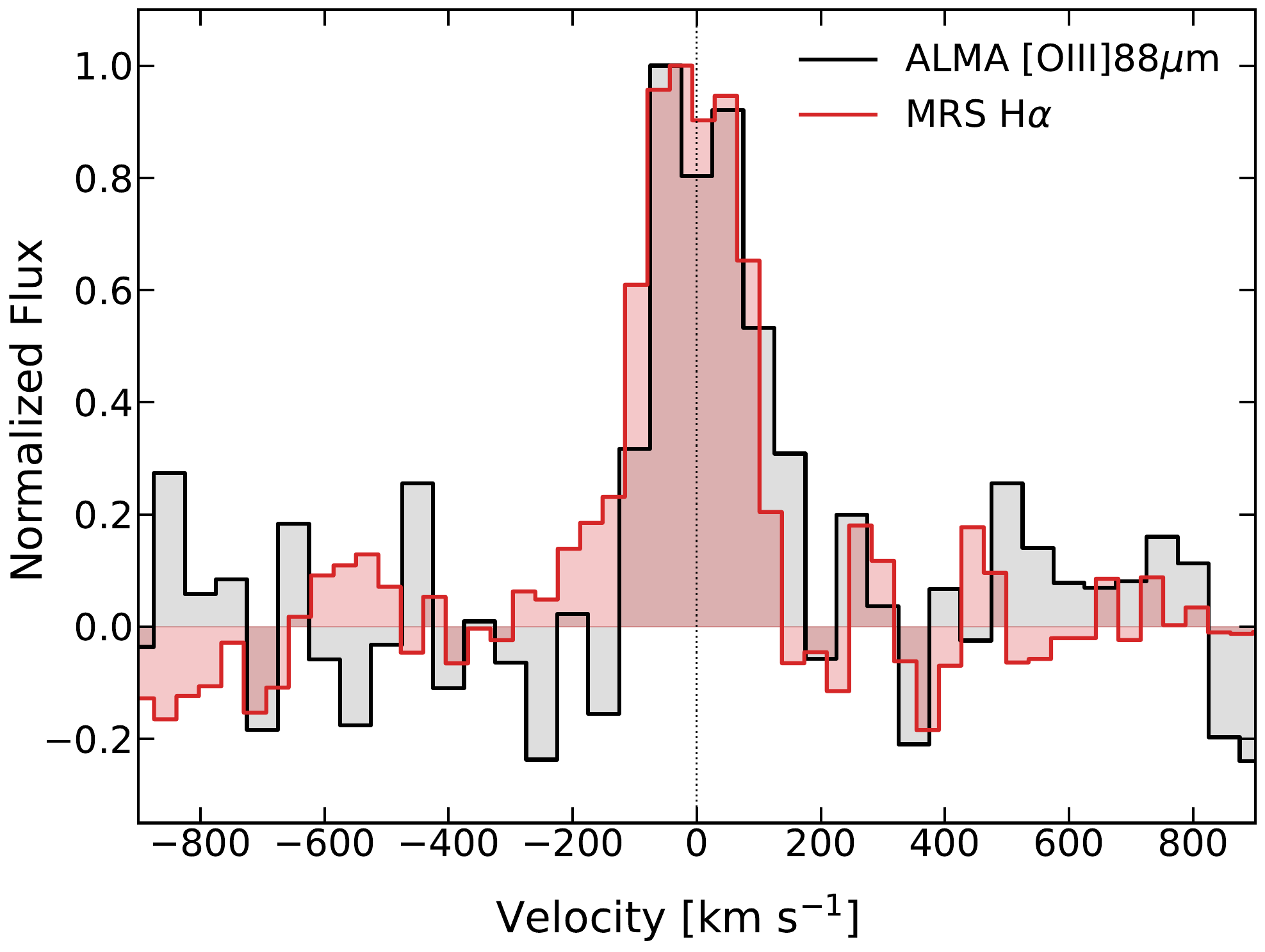}
      \caption{Comparison of the integrated H$\alpha$ emission line profile (red) with the ALMA spectrum of the [OIII]88$\mu$m line (black, \citealt{Tokuoka+22}). The systemic velocity corresponds to a redshift of 9.1096.}
         \label{fig:ALMA_comp}
\end{figure}

The overall kinematics of the ionized gas are traced by the profile of the integrated H$\alpha$ emission. The redshift, line profile, and velocity dispersion are consistent with the highly ionized gas traced by the far-IR [OIII]88$\mu$m emission measured with ALMA (see Figure \ref{fig:ALMA_comp}, \citealt{Hashimoto+18}, \citealt{Tokuoka+22}). The H$\alpha$ emitting gas is characterized by a velocity dispersion ($\sigma_\mathrm{V}$) of 69.2\,$\pm$\,5.5\,km\,s$^{-1}$, in agreement with the [OIII]88$\mu$m emission measured with ALMA ranging from 65.4\,$\pm$\,16.6\,km\,s$^{-1}$ \citep{Hashimoto+18} to 72.7\,$\pm$\,8.1\,km\,s$^{-1}$ \citep{Tokuoka+22}. The redshifts measured by the H$\alpha$ ($z$\,=\,9.1092\,$\pm$\,0.0002) and the [OIII]88$\mu$m lines ($z$\,=\,9.1096\,$\pm$\,0.0006, \citealt{Hashimoto+18} and $z$\,=\,9.1111\,$\pm$\,0.0006, \citealt{Tokuoka+22}) are also consistent with each other.

The H$\alpha$ clumps have a slightly different velocity, with a relative shift (N versus S clump) in central velocities of $-$36$\pm$20\,km\,s$^{-1}$. This could be due to the velocity field of the rotating disk identified in the ALMA [OIII]88$\mu$m emission line map \citep{Tokuoka+22}. However, clumps N and S show very different kinematics characterized by velocity dispersions of 113\,$\pm$\,33\,km\,s$^{-1}$ and 56\,$\pm$\,4\,km\,s$^{-1}$ (i.e., a factor of two), likely indicating the presence of outflows and increased turbulence in clump N associated with the rest-frame UV clump (C1; \citealt{Bradac+23,Stiavelli+23}). A deeper analysis combining the ALMA [OIII]88$\mu$m and the mid-IR JWST spectroscopy will explore the overall velocity field and the nature of these kinematic differences between clumps S and N in a follow-up work.

Under the assumption that the spatially resolved [OIII]88$\mu$m kinematics are compatible with a dispersion-dominated or a slow rotational system ($V_\mathrm{rot}$/$\sigma_\mathrm{V}$\,=\,0.69$^{+0.73}_{-0.26}$, \citealt{Tokuoka+22}), an estimate of the dynamical mass can be given assuming the virial expression

\begin{equation}
M_\mathrm{dyn}(M_\mathrm{\odot})= K \times \frac{R_\mathrm{hm} \times \sigma_\mathrm{V}^2}{G},
\end{equation}

\noindent
where $G$ is the gravitational constant with a value of 4.3\,$\times$\,10$^{-3}$\,pc\,M$_{\odot}^{-1}$\,km$^2$\,s$^{-2}$ and $K$ is set to a value of six (see \citealt{Bellocchi+13} for a discussion about the range of values for different mass distributions). The velocity dispersion ($\sigma_\mathrm{V}$\,=\,69\,$\pm$\,5\,km\,s$^{-1}$) is derived from the H$\alpha$ line profile after deconvolution with the instrumental response. The half mass radius in parsecs is represented by $R_\mathrm{hm}$, here traced by the H$\alpha$ half-light radius ($R_\mathrm{e}$) after correction by the PSF and with magnification as indicated below.

The intrinsic size of the integrated H$\alpha$ emission is given as the radius of the circle with an area equal to the ellipse containing half the light of the lensed galaxy after deconvolution with the MRS PSF for channel 1LONG in quadrature:  

\begin{equation}
R_\mathrm{e}(\mathrm{pc}) = 4.522 \times (a \times b)^{0.5} \times \mu^{-0.5},
\end{equation}

\noindent
where $\mu$ is the lensing magnification, and $a$ and $b$ are the semi-major and semi-minor axes (in milli-arcseconds) of the ellipse enclosing half the light after deconvolution with the PSF. For an effective radius $R_\mathrm{e}$\,=\,332\,$\pm$\,54\,pc, we obtained a dynamical mass of 2.4\,$\pm$\,0.5\,$\times$\,10$^9$\,$M_{\odot}$. This value is within the range derived from the spatially resolved [OIII]88$\mu$m velocity field
(0.7\,$-$\,3.7\,$\times$\,10$^9$\,$M_{\odot}$, \citealt{Tokuoka+22}). However, the H$\alpha$ ionized gas and the stellar light show a different structure (see Figure~\ref{fig:HaLineMap} and NIRCam images in \citealt{Bradac+23}, \citealt{Stiavelli+23}). The rest-frame optical appears to be dominated by the southern component, while the H$\alpha$ is more extended. Following the same assumptions as above, we estimated the dynamical mass for clump S only. For a gas velocity dispersion $\sigma_\mathrm{V}$\,=\,56\,$\pm$\,4\,km\,s$^{-1}$ and upper limit for the effective radius of 211\,pc, the dynamical mass corresponds to an upper limit of (1.0\,$\pm$\,0.2)\,$\times$\,10$^9$\,$M_{\odot}$. This mass is still a factor of 10 larger than the one reported with recent JWST NIRCam imaging for this region \citep{Bradac+23}. 

The differences between the estimated dynamical and stellar masses are too large, even when considering the uncertainties in their derivation due to the assumed ages for the stellar populations and 
due to the assumption of virialization, mass distribution, and kinematics (see \citealt{Bellocchi+13} and \citealt{Tokuoka+22}). The mass difference can in part be explained as being due to the available amount of cold gas. The gas fraction in MACS1149-JD1 is estimated as 30$\%$ of the total mass from the spatially resolved [OIII]88$\mu$m velocity field \citep{Tokuoka+22}. Our measurements give only an upper size for the emitting regions. If the size is substantially smaller, the derived dynamical mass would be reduced by the same factor, and therefore the combined stellar and gas mass would come closer to the  value derived for the dynamical mass. Higher angular resolution velocity maps with NIRSpec/JWST and ALMA and imaging at wavelengths longer than 4.4 $\mu$m with MIRI/JWST are required to more precisely establish the dynamical and stellar mass structure in MACS1149-JD1.

\section{Summary}\label{Sec:conclusion_Summary}

We have presented the first-ever spatially resolved H$\alpha$ line map of a galaxy at a redshift above nine: the lensed galaxy MACS1149-JD1 at a redshift of $z$\,=\,9.11. The direct detection of H$\alpha$ with MIRI/JWST provides reliable measurements of the spatially resolved instantaneous SFR and the ionizing photon production efficiency (the latter were obtained by also using available UV imaging), gas kinematics, and dynamical mass estimates.

The H$\alpha$ emitting gas shows a structure dominated by two spatially resolved clumps, N and S, separated by 0.5\arcsec, similar to the rest-frame UV continuum. However, the flux contribution is different, with the N and S clumps carrying 42$\%$ and 58$\%$ of the H$\alpha$ flux but 21$\%$ and 79$\%$ in the UV, respectively.  

The SFR derived from the H$\alpha$ luminosity ranges from 3.2 to 5.3\,$M_{\odot}$\,yr$^{-1}$ for sub-solar to solar metallicity, respectively, which is in good agreement with the UV-based estimates for less than 10\,Myr old stellar populations and with the value based on the [CII]158$\mu$m luminosity. These results support the hypothesis that internal dust extinction is not relevant and that the star formation is unobscured in MACS1149-JD1. 

The ionizing photon production efficiency log($\zeta_\mathrm{ion}$)\,=\,25.55\,$\pm$\,0.03\,Hz\,erg$^{-1}$ is well above the canonical value of 25.2\,Hz\,erg$^{-1}$ but within the range recently measured in other galaxies above a redshift of seven. The photon production efficiency shows a substructure with a factor three of difference between clump S and N, with values log($\zeta_\mathrm{ion}$)\,=\,25.47\,$\pm$\,0.03 and 
\,25.91\,$\pm$\,0.09\,Hz\,erg$^{-1}$, respectively. These values are in the range measured in galaxies at redshifts above six by JWST, and they do not indicate any evidence of evolution with redshift. 

The integrated H$\alpha$ equivalent width measured in MACS1149-JD1 has a value of ({726$^{+660}_{-182}$\,$\AA$}), which is a factor of two less than the extrapolated value at redshift 9.11 derived from samples of galaxies up to redshift eight \citep{Rinaldi+23}. Clumps S and N show a large difference in their equivalent widths ({531$^{+300}_{-96}\AA$} and $\geq$\,{1951\,$\AA$}), indicating the presence of different stellar populations in these two regions. The large value in clump N indicates a very young ({<5 Myr}) stellar burst, while the intermediate value in clump S is more consistent with a star formation over a longer period of time ({$\sim$50\,Myr}). The EW$_0$\,(H$\alpha$) for MACS1149-JD1 indicates that the stellar mass is dominated by clump S, while clump N appears to be a recent burst in the galaxy. The positions of MACS1149-JD1 and the clumps in the log($\zeta_\mathrm{ion}$)\,$-$\,EW$_{0}$\,(H$\alpha$) plane reflect the substructure in the stellar populations.  

The overall H$\alpha$ emitting gas kinematics (redshift of 9.1092 and velocity dispersion of 69\,$\pm$\,5\,km\,s$^{-1}$)  agree with that of the [OIII]88$\mu$m line previously measured with ALMA. The dynamical mass derived from the profile of the H$\alpha$ line and the size of the H$\alpha$ surface brightness corresponds to (2.4\,$\pm$\,0.5)\,$\times$\,10$^9$\,$M_{\odot}$. This mass is within the range measured from spatially resolved [OIII]88$\mu$m emission \citep{Tokuoka+22}. The velocity dispersion of the H$\alpha$ emitting gas in clump N (113\,$\pm$\,33\,km\,s$^{-1}$) is a factor of two higher than in clump S, likely tracing the presence of a local outflow associated with the UV-bright {clump}.

\begin{acknowledgements}
The  authors  thank  Akio Inoue and Takuya Hashimoto for sharing the ALMA spectrum of [OIII]88$\mu$m emission line. J.A-M., L.C., A.C-G. acknowledge support by grant PIB2021-127718NB-100, A.A-H. by grant PID2021-124665NB-I00 from the Spanish Ministry of Science and Innovation/State Agency of Research MCIN/AEI/10.13039/501100011033 and by “ERDF A way of making Europe”. S.E.I.B. is supported by the Deutsche Forschungsgemeinschaft (DFG) through Emmy Noether grant number BO 5771/1-1. J.M., A.B., and G.O. acknowledges support from the Swedish National Space Administration (SNSA). K.I.C. and EI acknowledge funding from the Netherlands Research School for Astronomy (NOVA). K.I.C. acknowledges funding from the Dutch Research Council (NWO) through the award of the Vici Grant VI.C.212.036. S.G. acknowledges financial support from the Villum Young Investigator grant 37440 and 13160. J.H. and D.L. were supported by a VILLUM FONDEN Investigator grant (project number 16599). O.I. acknowledges the funding of the French Agence Nationale de la Recherche for the project iMAGE (grant ANR-22-CE31-0007). R.A.M. acknowledges support from the Swiss National Science Foundation (SNSF) through project grant 200020\_207349. J.P.P. acknowledges financial support from the UK Science and Technology Facilities Council, and the UK Space Agency. LB and F.W. acknowledge support from the ERC Advanced Grant 740246 (Cosmic\_Gas). A.E. and F.P. acknowledge support through the German Space Agency DLR 50OS1501 and DLR 50OS2001 from 2015 to 2023. GP-G acknowledges support from grants PGC2018-093499-B-I00 and PID2022-139567NB-I00 funded by Spanish Ministerio de Ciencia e Innovación MCIN/AEI/10.13039/501100011033, FEDER, UE.
      
The work presented is the effort of the entire MIRI team and the enthusiasm within the MIRI partnership is a significant factor in its success. MIRI draws on the scientific and technical expertise of the following organisations: Ames Research Center, USA; Airbus Defence and Space, UK; CEA-Irfu, Saclay, France; Centre Spatial de Liége, Belgium; Consejo Superior de Investigaciones Científicas, Spain; Carl Zeiss Optronics, Germany; Chalmers University of Technology, Sweden; Danish Space Research Institute, Denmark; Dublin Institute for Advanced Studies, Ireland; European Space Agency, Netherlands; ETCA, Belgium; ETH Zurich, Switzerland; Goddard Space Flight Center, USA; Institute d'Astrophysique Spatiale, France; Instituto Nacional de Técnica Aeroespacial, Spain; Institute for Astronomy, Edinburgh, UK; Jet Propulsion Laboratory, USA; Laboratoire d'Astrophysique de Marseille (LAM), France; Leiden University, Netherlands; Lockheed Advanced Technology Center (USA); NOVA Opt-IR group at Dwingeloo, Netherlands; Northrop Grumman, USA; Max-Planck Institut für Astronomie (MPIA), Heidelberg, Germany; Laboratoire d’Etudes Spatiales et d'Instrumentation en Astrophysique (LESIA), France; Paul Scherrer Institut, Switzerland; Raytheon Vision Systems, USA; RUAG Aerospace, Switzerland; Rutherford Appleton Laboratory (RAL Space), UK; Space Telescope Science Institute, USA; Toegepast- Natuurwetenschappelijk Onderzoek (TNO-TPD), Netherlands; UK Astronomy Technology Centre, UK; University College London, UK; University of Amsterdam, Netherlands; University of Arizona, USA; University of Cardiff, UK; University of Cologne, Germany; University of Ghent; University of Groningen, Netherlands; University of Leicester, UK; University of Leuven, Belgium; University of Stockholm, Sweden; Utah State University, USA. A portion of this work was carried out at the Jet Propulsion Laboratory, California Institute of Technology, under a contract with the National Aeronautics and Space Administration. We would like to thank the following National and International Funding Agencies for their support of the MIRI development: NASA; ESA; Belgian Science Policy Office; Centre Nationale D'Etudes Spatiales (CNES); Danish National Space Centre; Deutsches Zentrum fur Luft-und Raumfahrt (DLR); Enterprise Ireland; Ministerio De Econom\'ia y Competitividad; Netherlands Research School for Astronomy (NOVA); Netherlands Organisation for Scientific Research (NWO); Science and Technology Facilities Council; Swiss Space Office; Swedish National Space Board; UK Space Agency. 

This work is based on observations made with the NASA/ESA/CSA James Webb Space Telescope. The data were obtained from the Mikulski Archive for Space Telescopes at the Space Telescope Science Institute, which is operated by the Association of Universities for Research in Astronomy, Inc., under NASA contract NAS 5-03127 for \textit{JWST}; and from the \href{https://jwst.esac.esa.int/archive/}{European \textit{JWST} archive (e\textit{JWST})} operated by the ESDC.

This paper makes use of the following ALMA data: ADS/JAO.ALMA\#2016.1.01293.S and ADS/JAO.ALMA\#2017.1.01493.S. ALMA is a partnership of ESO (representing its member states), NSF (USA) and NINS (Japan), together with NRC (Canada), MOST and ASIAA (Taiwan), and KASI (Republic of Korea), in cooperation with the Republic of Chile. The Joint ALMA Observatory is operated by ESO, AUI/NRAO and NAOJ. The National Radio Astronomy Observatory is a facility of the National Science Foundation operated under cooperative agreement by Associated Universities, Inc. 

This research made use of Photutils, an Astropy package for detection and photometry of astronomical sources \citep{larry_bradley_2022_6825092}.

\end{acknowledgements}

\bibliographystyle{aa} 
\bibliography{bibliography.bib} 

\end{document}